\documentclass[prl,final,aps,twocolumn]{revtex4-1}
\usepackage{amsmath,appendix}
\usepackage[usenames,dvipsnames]{color}
\usepackage{graphicx,textcase} 
\usepackage{cancel}
\usepackage{ulem}

\begin{document}

\title{ 
Chiral Spin and Orbital Angular Momentum Textures in Mn Chains on W(110): Interplay of Spin-Orbit Coupling and Crystal Field Effects
}


\author{M. M. Bezerra-Neto$^{1}$}
\author{Y. O. Kvashnin$^{2}$}
\author{A. Bergman$^{2}$}
\author{R. Cardias$^{3}$}
\author{R. B. Muniz$^{3}$}
\author{O. Eriksson$^{2,4}$}
\author{M. I. Katsnelson$^{5}$}
\author{A. B. Klautau$^{6}$}

\affiliation{$^{1}$Instituto de Engenharia e Geoci\^encias, Universidade Federal do Oeste do Par\'a, Santar\'em, PA, Brazil}
\affiliation{$^{2}$Uppsala University, Department of Physics and Astronomy, Box 516, SE-75120, Uppsala, Sweden}
\affiliation{$^{3}$Instituto de F\'\i sica, Universidade Federal Fluminense, 24210-346 Niter\'oi, Rio de Janeiro, Brazil}
\affiliation{$^{4}$Wallenberg Initiative Materials Science for Sustainability (WISE), Uppsala University, Box 516, SE-75120, Uppsala, Sweden}
\affiliation{$^{5}$Institute for Molecules and Materials, Radboud University, Heijendaalseweg 135, 6525AJ, Nijmegen, Netherlands}
\affiliation{$^{6}$Faculdade de F\'\i sica, Universidade Federal do Par\'a, Bel\'em, PA, Brazil}

\date{\today}

\begin{abstract}

\noindent

Stabilization of unusual spin-orbit driven magnetic orderings are achieved for chains of Mn atoms deposited on a W(110) substrate. First-principles electronic structure calculations show that the ground state spin configuration is non-collinear, forming  
chiral spiral-like structures,
driven by competing nearest and next-nearest neighbor interactions. 
The orbital magnetic moments are also found to exhibit non-collinear ordering that, interestingly,   
tend to align in-plane for some systems with an orientation distinctly differently from that of the spin moment.
We analyse the mechanism behind such 
behaviour, and find that it is due to the competition between the spin-orbit interaction and crystal-field splitting.
Model calculations based on this assumption reproduce the main findings observed in our first-principles calculations.

\end{abstract}

\maketitle

\section{Introduction}
 
Chiral spin textures, such as skyrmions  \cite{fert2013skyrmions,romming2013writing} and spin spirals 
\cite{yoshida2012conical,bode2007chiral,menzel2012information,steinbrecher2018non}, 
 have attracted intense interest not only because of the fundamental physics governing their origins but also due to their potential applications in novel information storage technologies.
Ultrathin magnetic films deposited on heavy metal substrates are good candidates for displaying complex spin textures. 
In general, the chiral spin states observed in these systems are induced by an interplay between the Dzyaloshinskii-Moriya (DM) interaction 
\cite{dzyaloshinskyThermodynamicTheoryWeak1958,moriyaAnisotropicSuperexchangeInteraction1960}, the bilinear Heisenberg  exchange interaction, 
and the magnetocrystalline anisotropy.
Monolayers of Mn on W(110) \cite{yoshida2012conical,bode2007chiral} 
 and W(001) \cite{mnw0012023} 
represent such cases where the combination of the relatively strong spin-orbit interaction of tungsten and the broken inversion symmetry at the interfaces results in a significant interface-induced DM interaction. As a result, the ground-state spin magnetic moments display different types of spin spirals, depending on the number of Mn atomic planes 
and the surface plane direction. 
The spin structure of a Mn trilayer on the W(001) surface is a  
spin spiral that propagates along the [110] direction, with an angle close to $90^{o}$ between magnetic moments of adjacent Mn rows \cite{mnw0012023}.  
For a single Mn overlayer on W(110) the spin magnetic moments exhibit a cycloidal magnetic order as observed in Ref. \onlinecite{bode2007chiral}. The Mn bilayer show a conical spin-spiral state along the [001] direction \cite{yoshida2012conical} and in both cases a unique rotational sense is enforced by the chiral DM interaction.


With the advent of {\it orbitronics} \cite{go2018intrinsic,Go_2021,Go2020,Go2023Nature,Mankovsky2024,fert2024,lee2024electric}, which 
exploits the transport of orbital angular momentum through materials via orbital currents, and 
 its influence on spin degrees of freedom 
by means of the spin-orbit interaction,
the development of model Hamiltonians has gained momentum in recent years, and several new approaches have been discussed in the literature \cite{SECCHI201561,mankovsky2020extension,laszloffy2019magnetic,grytsiuk2020topological,cardiasDzyaloshinskiiMoriyaInteractionAbsence2020}.
 %
 In this scenario, orbital effects became a focus of studies and have been explored in various aspects. These include the orbital Hall effect, where an external electric field induces an orbital angular momentum current that flows in the transverse direction of the applied field \cite{tanaka2008intrinsic,kontani2009giant,go2018intrinsic}; their role in magnons that besides spin may also carry orbital angular moment  \cite{neumann2020orbital};
 as well as the capability of orbital angular momentum to exert torque on magnetic textures \cite{PhysRevResearch.2.013177}. 
Furthermore, the study of orbital moments can be pivotal in the formation of Rashba-type surface band splittings, which may lead to chiral orbital angular momentum states \cite{park2011orbital,park2013orbital}, and in the connection between the orbital magnetic moment anisotropy and the DM interaction \cite{kim2018correlation}, among others. 
  Moreover, the non-collinearity of spin and orbital moments has been discussed for bulk materials in the context of weak ferromagnetism \cite{Sandratskii1,Sandratskii2} and for U compounds \cite{Sandratskii3}, as well as for deposited quasi-one-dimensional chains on step edge surfaces in the case of a collinear ferromagnetic spin arrangement \cite{PhysRevB.69.212410}. 
 Nevertheless, the dependence of the magnetic ordering of the spin and orbital moments on the width of adsorbed chains has not been explored. 
To obtain a realistic model Hamiltonian based on electronic structure, which takes all relevant interactions into account in a general way,  
remains a considerable challenge. One of the promising approaches is based on the consideration of small rotations near a given magnetic configuration (``magnetic force theorem'') as was previously suggested for purely spin rotations \cite{liechtensteinLocalSpinDensity1987,PhysRevB.61.8906}. Its generalization for the systems with unquenched orbital moments \cite{SECCHI201561,PhysRevB.94.115117} allows to write explicit formulas for spin-spin, orbital-orbital, and spin-orbital interactions but until now they were not applied to specific real materials.  
 
Here, by means of first-principles calculations, we demonstrate the occurrence of atomic scale  cycloidal-type spin-spiral magnetic structures,  
 with antiferromagnetic short-range order, 
 for Mn chains placed on top of a W(110) surface.
 It is noteworthy that we are dealing with Mn chains comprising relatively small number of atoms. Thus, the magnetic structures discussed here are not spin spirals in the traditional sense, since this terminology is typically used for infinite systems. Nevertheless, we here refer to them as spiral-like structures, since the 
 cycloid 
pitch far away from the edges approaches that of the regular spin spiral of an infinite system.


The choice of these systems was motivated by the remarkable magnetic properties observed for Mn overlayers on W(110) \cite{yoshida2012conical,bode2007chiral}, as well as by the possibility of fabricating atomic chains of transition metal atoms on heavy metal substrates.
We have examined the ground state spin and orbital magnetic configurations of chains and nanoribbons (stripes) of Mn deposited along two distinct crystallographic directions of a W(110) surface. 
We found that the ground-state spin configuration of the Mn nanostructure changes with the increase of its  width.  
 The mono-atomic Mn chain 
 exhibits a spin-spiral-like configuration with cycloidal nature and almost antiferromagnetic (AFM) ordering between nearest neighbors along the chain, 
whereas 
Mn ribbons with three atoms in width 
 display AFM order between rows, yielding a more complex, multi-row cycloidal spin-spiral-like  structure state. 
%
Such chiral magnetic orders are also revealed for the orbital moments in some cases, and depending on the length and width of the Mn nanostructures, the spin and orbital spiral configurations may exhibit different patterns. 
We discuss the obtained magnetic chirality in terms of 
effective low energy Hamiltonians that contain spin-spin, orbital-orbital as well as as spin-orbital exchange.
The exotic behaviour concerning the  
 distinguishing 
orbital spiral-like configurations with respect to the spin arrangement is microscopically attributed to a competition between crystal field and spin-orbit coupling, which are comparable in this material. Our model calculations, based on this assumption, reproduce the main findings observed in the first-principles calculations.

\section{Computational details}
The density functional theory based calculations 
were performed using the first principles real-space linear muffin-tin orbital method within the atomic sphere approximation (RS-LMTO-ASA)  \cite{frota-pessoaFirstprinciplesRealspaceLinearmuffintinorbital1992,klautauOrbitalMoments3d2005,bergmanMagneticInteractionsMn2006,bergmanMagneticStructuresSmall2007,bezerra-netoComplexMagneticStructure2013,klautauMagneticPropertiesCo2004,frota-pessoaInfluenceInterfaceMixing2002,bergmanNoncollinearMagnetisationClusters2006,PhysRevB.85.014436,PhysRevB.93.014438,PhysRevB.105.224413,PhysRevB.103.L220405},  which has been generalized to treat non-collinear magnetism  \cite{bergmanNoncollinearMagnetisationClusters2006,cardias2020first,cardiasDzyaloshinskiiMoriyaInteractionAbsence2020,carvalho2023correlation,PhysRevB.108.224408}. 
This method is based on the LMTO-ASA formalism  \cite{andersenLinearMethodsBand1975} and employs the recursion method \cite{haydockRecursiveSolutionSchrodinger1980} to solve the eigenvalue problem directly in real space.   
All calculations are fully self-consistent and exchange-correlation effects were taken into account within 
the local spin density approximation (LSDA)  \cite{barthLocalExchangecorrelationPotential1972}.
 For the collinear and the non-collinear calculations, we included the spin-orbit coupling term, $\lambda L\cdot S$,  self-consistently at each variational step, where $L$ and $S$ represent the orbital and spin angular momentum operators, and $\lambda$ is the calculated spin-orbit coupling strength \cite{PhysRevB.42.2707,PhysRevB.69.104401}.  
 A detailed description of the 
method for the situation of the non-collinear calculations can be found on  Ref.~\cite{bergmanNoncollinearMagnetisationClusters2006}.
 In order to minimize the risk of finding magnetic orderings that correspond to only a local minimum, several starting guesses were used for each system \cite{bergmanNoncollinearMagnetisationClusters2006,cardias2020first,cardiasDzyaloshinskiiMoriyaInteractionAbsence2020,carvalho2023correlation,PhysRevB.108.224408}. 

We have considered Mn nanostructures, with different geometries, placed atop 
a W(110) surface.
Since we use a real-space method, the W(110) substrate is modeled by a cluster containing approximately 8000 atoms positioned in a bcc lattice with the experimental lattice parameter of W. In order to simulate the vacuum outside the bcc surface and to treat charge transfers correctly, we included two overlayers of empty spheres above the W surface. 
The calculations of the Mn adsorbed nanochains have been performed by embedding the clusters as a perturbation on the previously self-consistently converged W(110) surface.
In order to achieve an accurate description of the hybridization with the substrate, the Mn sites and the closest shells of neighboring atoms (or empty spheres) around the Mn nanostructures were recalculated self-consistently. 
The continued fraction in the recursion method was appended with the Beer-Pettifor terminator \cite{beerRecursionMethodEstimation1984} after 30 recursion levels. 
Structural relaxations have not been included in most of the systems considered 
here. For the Mn nanowire (NW) on W(110), displayed in  Fig.~\ref{fig:systems}, we have performed electronic structure calculations considering an inward perpendicular relation towards the W(110) surface of 4\%, and found that the difference between the magnetic moments values with respect to the unrelaxed case is $\approx 8\%$, whereas the Mn orbital magnetic moments change by less than 0.005 $\mu_{B}$. 
%
The $J_{ij}$ and $\vec{D}_{ij}$ parameters between Mn atoms (values not shown), on the other hand, are affected by such relaxations, but 
no sign changes were observed. The angles between different moments of the NW change only by a few degrees, and the spin-spiral-like pattern is maintained, even when a relaxation of 4\% is considered. The effective magnetic interactions between Mn atoms are much larger than the Mn-W ones.

The Heisenberg bilinear exchange interactions $J_{ij}$ and the DM vectors $\vec{D}_{ij}$ were calculated using the Liechtenstein-Katsnelson-Antropov-Gubanov (LKAG) formula \cite{liechtensteinLocalSpinDensity1987,PhysRevB.61.8906},
as  implemented in the RS-LMTO-ASA method \cite{frota2000exchange,szilvainteratomicExchangeInteractions2013,szilvaTheoryNoncollinearInteractions2017,cardias2020first,cardiasDzyaloshinskiiMoriyaInteractionAbsence2020}. 
Both $J_{ij}$ and $\vec{D}_{ij}$ have been obtained from the ferromagnetic configuration of the nanoclusters.
Their values are then used to analyze
the spin magnetic orderings explored in the full non-collinear calculations. 
%


\par
\section{Results and Discussions}

\begin{figure}[htp]
\centering
\includegraphics[scale=0.35]{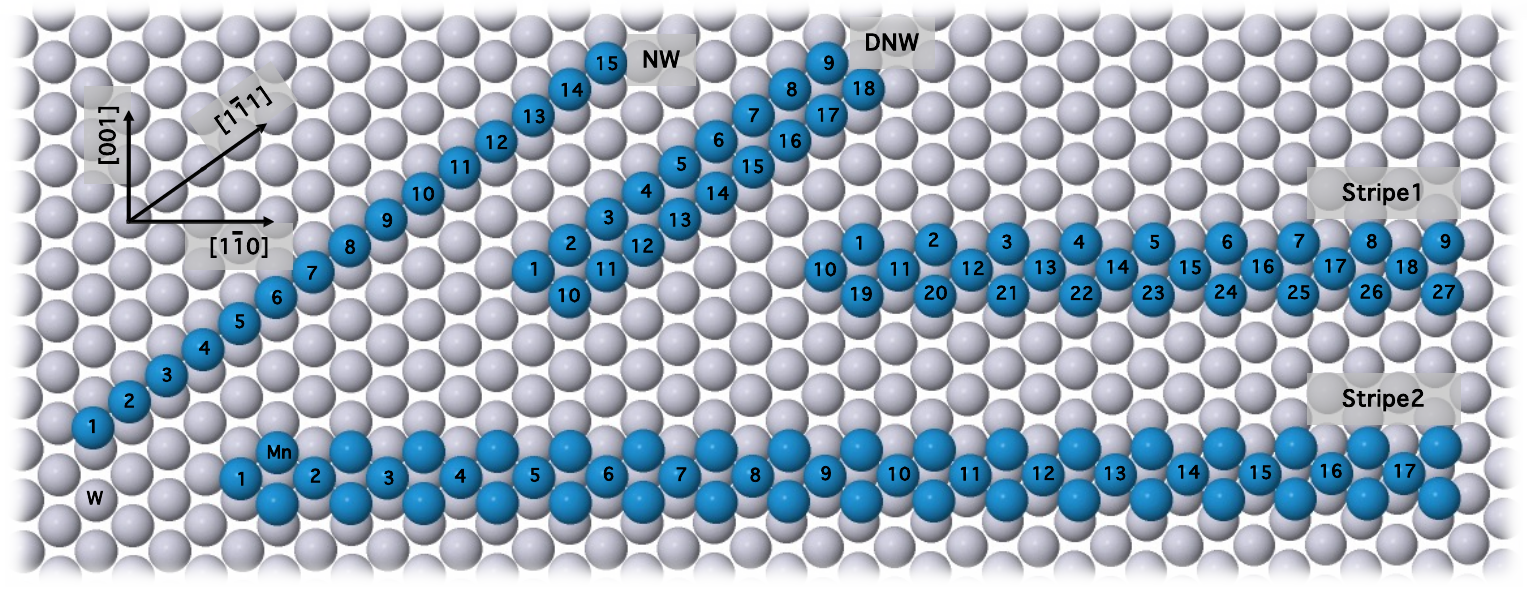}
\caption{
(Color online) Mn nanostructures adsorbed on a W(110) surface:
linear nanowire (NW), double nanowires (DNW), nanostripes (Stripe-1 and Stripe-2).
The blue (dark) and gray (light) balls indicate Mn and W atoms, respectively.}
\label{fig:systems}
\end{figure}

Fig.~\ref{fig:systems} illustrates the finite Mn nanostructures adsorbed on W(110) that were investigated here. First, 
we examined
the ground-state spin magnetic configurations of these nanostructures and, subsequently, their orbital magnetic orderings. 

We begin the discussion with a monoatomic chain, hereafter referred to as NW (short for nanowire) with 3.8 nm in length, containing 15 Mn atoms arranged along the [$1\overline{1}1$] direction. Next, we inspect a double nanowire (DNW), composed of two Mn chains that are placed side by side along the [$1\overline{1}1$] direction, both of which are 2.2 nm long and contain 9 atoms each. Finally, we study two finite Mn stripes of different lengths, both three atoms wide and deposited along the [$1\overline{1}0$] direction. They are referred to as Stripe-1 and Stripe-2, and are 3.58 nm and 7.17 nm long, respectively.

It 
is noteworthy
that the spin magnetic moments of the Mn atoms quite generally depend upon the number of nearest neighbors. Fewer neighbors yield higher spin magnetic 
moments, with values ranging from 3.98 $\mu_\text{B}$/atom, for atoms located at the NW edges, to 3.08 $\mu_\text{B}$/atom for the most central Mn site on Stripe-2 (for details, see Fig.~\ref{fig:angle-ms-mo-nw}). It is also found here that the Mn atoms induce a relatively small spin polarization on the W atoms adjacent to the Mn nanostructures, with values  $\sim$ 0.20 $\mu_\text{B}$/atom. 

\begin{figure}[htp]
\centering
\includegraphics[scale=0.43]{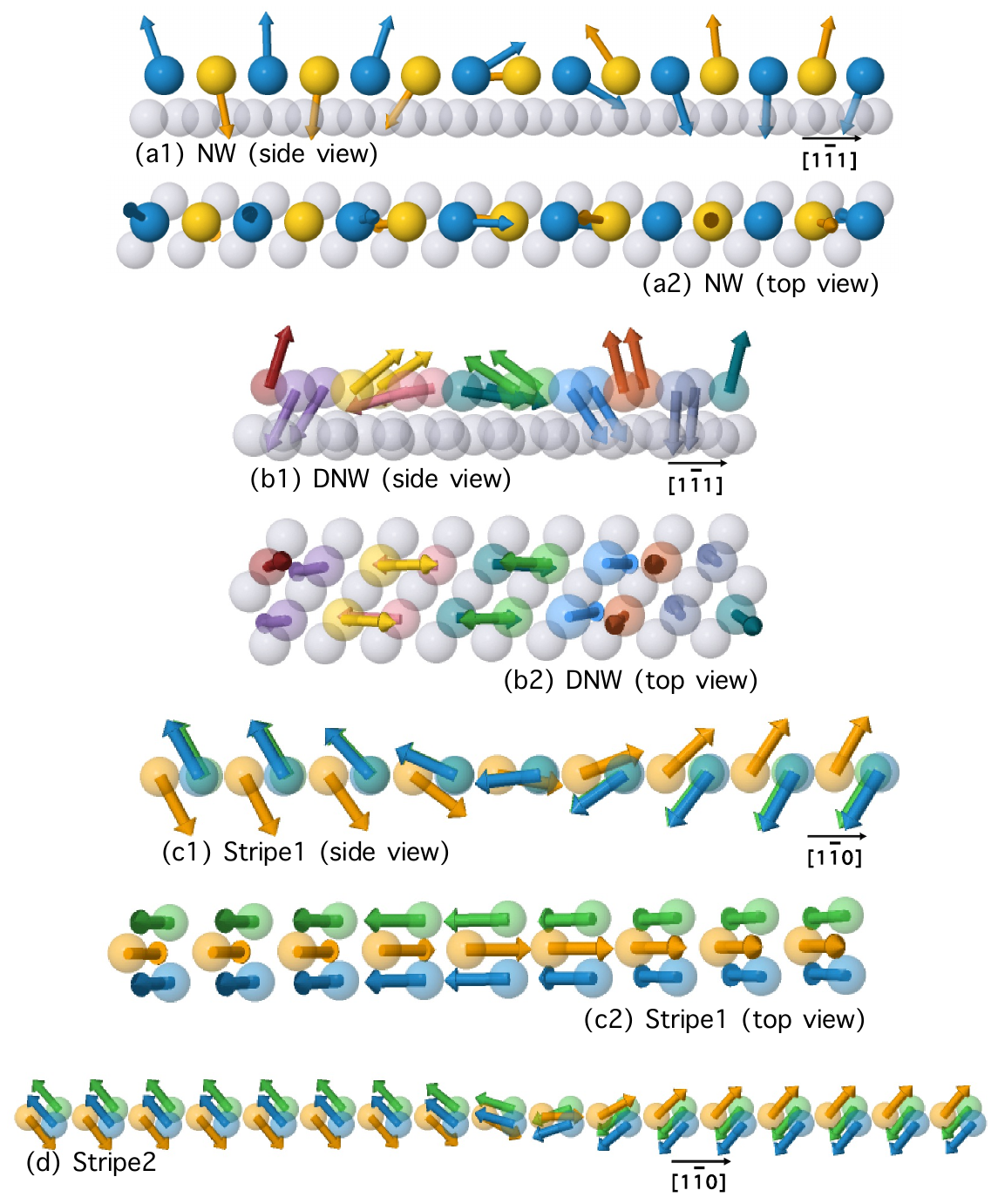}
\caption{
Non-collinear ground-state spin magnetic configurations (moments shown by arrows) for Mn nanostructures adsorbed on a W(110) surface. (a1,a2) Nanowire (NW) with 15 Mn atoms; (b1,b2) Double nanowire (DNW) with 9 Mn atoms in length; (c1,c2) Stripe-1 with with 3 Mn atoms in breadth and 9 Mn atoms in length; (d) Stripe-2 with with 3 Mn atoms in breadth and 17 Mn atoms in length.
}
\label{fig:spin-conf}
\end{figure}

Fig.~\ref{fig:spin-conf} displays the non-collinear ground-state magnetic spin configurations calculated self-consistently for the nanostructures described above.
The ground-state spin moments of the NW exhibits a 
%
cycloidal spin spiral-type arrangement along the [$1\overline{1}1$] direction, with a half periodicity of $\sim$ 2.7 nm. The angles between spin magnetic moments of nearest neighbors (NN) Mn atoms (central sites) are $\sim$ 140$^{\circ}$, and the spin chirality is clockwise (see Fig.~\ref{fig:colormap-angnw} for details). 
%
%
The spin texture in the DNW shows a complex chiral non-collinear arrangement where the spin moments form a cycloidal spiral pattern along the [$1\overline{1}1$] direction. Within each line, the spin magnetic moments of nearest neighbor Mn atoms tend to be oriented antiparallel to each other, 
with the moments showing a slight variation in angles, clustering $\sim$~100$^{\circ}$ (e.g. central pairs denoted by (5,6) and (13,14) in Fig.~\ref{fig:systems}) to $\sim$~150$^{\circ}$ (e.g. edge pairs (10,11) and (7,8)).
The spin magnetic moments of next nearest neighbor Mn atoms located in different lines are almost parallel (e.g. pairs labeled (2,10) and (6,14) in Fig.~\ref{fig:systems}, as well as those depicted by same colors in Fig.~\ref{fig:spin-conf} (b1) and (b2)). 
%
The first-principles calculations for the nano-stripes also reveal cycloidal spin spiral-like ground states with long-range modulations along the [$1\overline{1}0$] direction. Fig.~\ref{fig:spin-conf} shows that nearest-neighbor antiferromagnetic interactions between Mn atoms in adjacent rows play a dominant role in these systems. However, there is also a superimposed modulation along the [$1\overline{1}0$] direction that leads to noncollinear magnetic structures with half-periods of $\sim$4 nm and $\sim$7 nm for Stripe-1 and Stripe-2, respectively (only half-periods fit within their lengths). These spin configurations resemble those observed by spin-polarized scanning tunneling microscopy measurements on a Mn/W(110) monolayer, albeit with a somewhat different spin spiral period of $\sim$6 nm along the [$1\overline{1}0$] direction \cite{bode2007chiral}.

%

%

\begin{figure}[htp]
\includegraphics[scale=0.42]{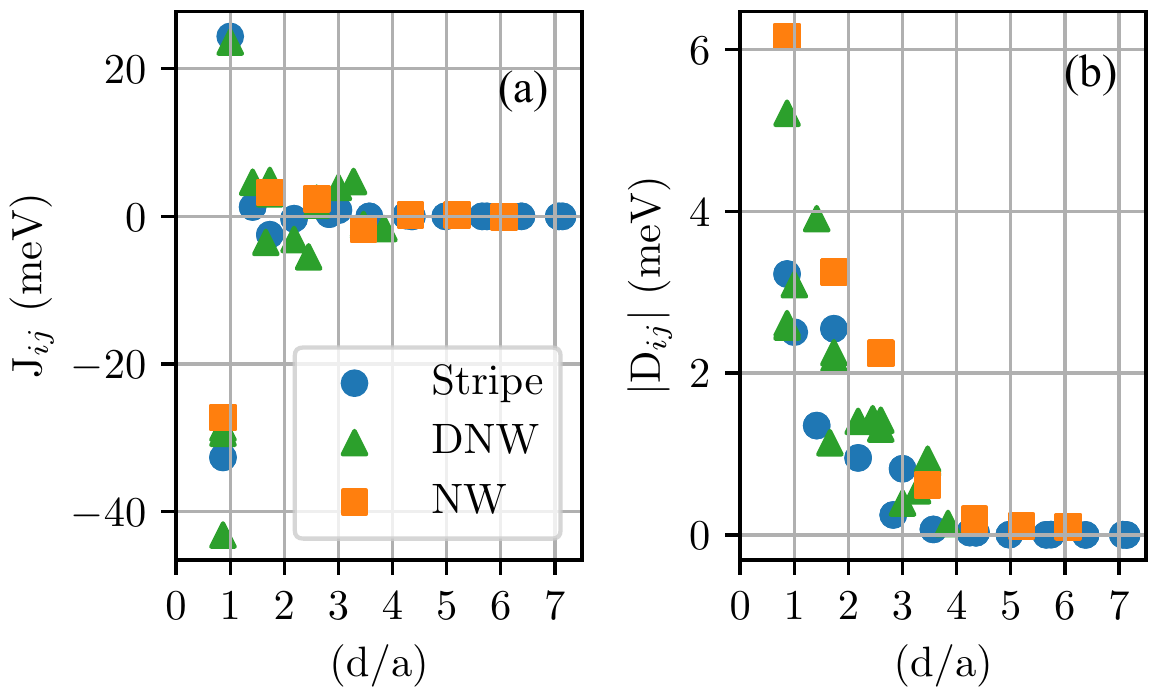}
\caption{Mn-Mn (a)  bi-linear Heisenberg exchange ($J_{ij}$) and (b) the Dzyaloshinskii-Moriya strengths ($|\vec{D}_{ij}|$) interactions calculated for the Mn nanostructures in their ferromagnetic configurations. The interactions between Mn atoms around the nanostructure central regions are plotted as functions of the inter atomic Mn distances: orange squares refer to the nanowire (NW); green triangles  are for the double-nanowire (DNW); and blue circles  stand for the Stripe-2. 
}
\label{fig:jij-dm}
\end{figure}

The calculated Heisenberg exchange interactions and the modulus of the DM vectors for the Mn nanostructures are shown in Fig.~\ref{fig:jij-dm} for several pairs of Mn atoms as functions of their interatomic distances.
We have only included the values for pairs of Mn atoms located around the center of the nanostructures. The results for atoms near the ends of the rows do not differ much from those illustrated in Fig.~\ref{fig:jij-dm} and basically show the same trends.
It is clear from Fig.~\ref{fig:jij-dm} that the bilinear exchange interactions $J_{ij}$ for the NW, DNW and Stripe-1 nanostructures decay in an oscillatory manner and have finite but significant ranges.
We note that the $|J_{ij}|$ for the NW is larger for nearest-neighbors Mn atoms and decreases relatively fast with the interatomic distance. Similar trends of the distance dependence of the parameters $|J_{ij}|$ are also found for all systems investigated here, as illustrated in Fig.~\ref{fig:jij-dm}(a) . 
It should be noted that positive (negative) values of $J_{ij}$ represent effective ferromagnetic (antiferromagnetic) couplings between spin magnetic moments located at sites $i$ and $j$.

%
%

%
 As shown in Fig.~\ref{fig:jij-dm}(b) the strength of the DM vector ($|\vec{D}_{ij}|$) associated with Mn NN pairs is much smaller than the corresponding values of $|J_{ij}|$ and the same happens for more distant neighbors. The DM vectors are finite and interactions between close-neighboring atoms are the most relevant. Figs.~\ref{fig:dm-direction}(a-c) show the calculated directions of the DM vectors associated with a reference atom (highlighted in green) and its neighbors for the NW, DNW and Stripe-1 systems, respectively. 
For the NW, the DM vectors between the NN pairs of Mn atoms are along the $[\overline{1}12]$ direction, i.e., they are in-plane and perpendicular to the chain direction (see Fig.~\ref{fig:dm-direction}(a) of the appendix). 
For stripes, the values of $|\vec{D}_{ij}|$ and $|J_{ij}|$ are much closer than in the NW case, as a small out-of-plane component of $\vec{D}_{ij}$ appears, but when $\vec{D}_{ij}$ is summed over all NN $j$ atoms, the net out-of-plane component nearly vanishes, and the total DM vector is practically in-plane and perpendicular to the length of the stripe. The same applies to the DNW. This specific direction of the DMI favors an arrangement of tilted spin magnetic moments along the length of the Mn nanostructures, and the balance between $J_{ij}$ and $\vec{D}_{ij}$ explains the ground-state spin-spiral like magnetic configurations found in these systems.
%
%
%

%
 In order to investigate the influence of the magnetic anisotropy on the magnetic configuration of these systems, we have also performed self-consistent calculations considering a collinear antiferromagnetic  ordering for the NW, and the magnetic anisotropy energy (MAE)  was obtained by calculating the energy difference ($\Delta {E}$) between the cases where the Mn moments are perpendicular to the plane (${E}_{\perp}$), in-plane (${E}_{\parallel}$) along the wire direction,  as well as in-plane and perpendicular to the wire direction  (${E}_{[\overline{1}12]}$). We obtained $\Delta {E}={E}_{[\overline{1}12]} - {E}_{\parallel}$ $\sim$ 0.2 meV and $\Delta {E}={E}_{\perp} - {E}_{\parallel}$ $\sim$ 1 meV, indicating that the easy axis is along the [$1\overline{1}1$] direction (in the plane). 
 The MAE in these systems is smaller than the NN DM and Heisenberg interactions, a finding which is rather common for transition metals. 
 %

\begin{figure}[htp]
\centering
\includegraphics[scale=0.43]{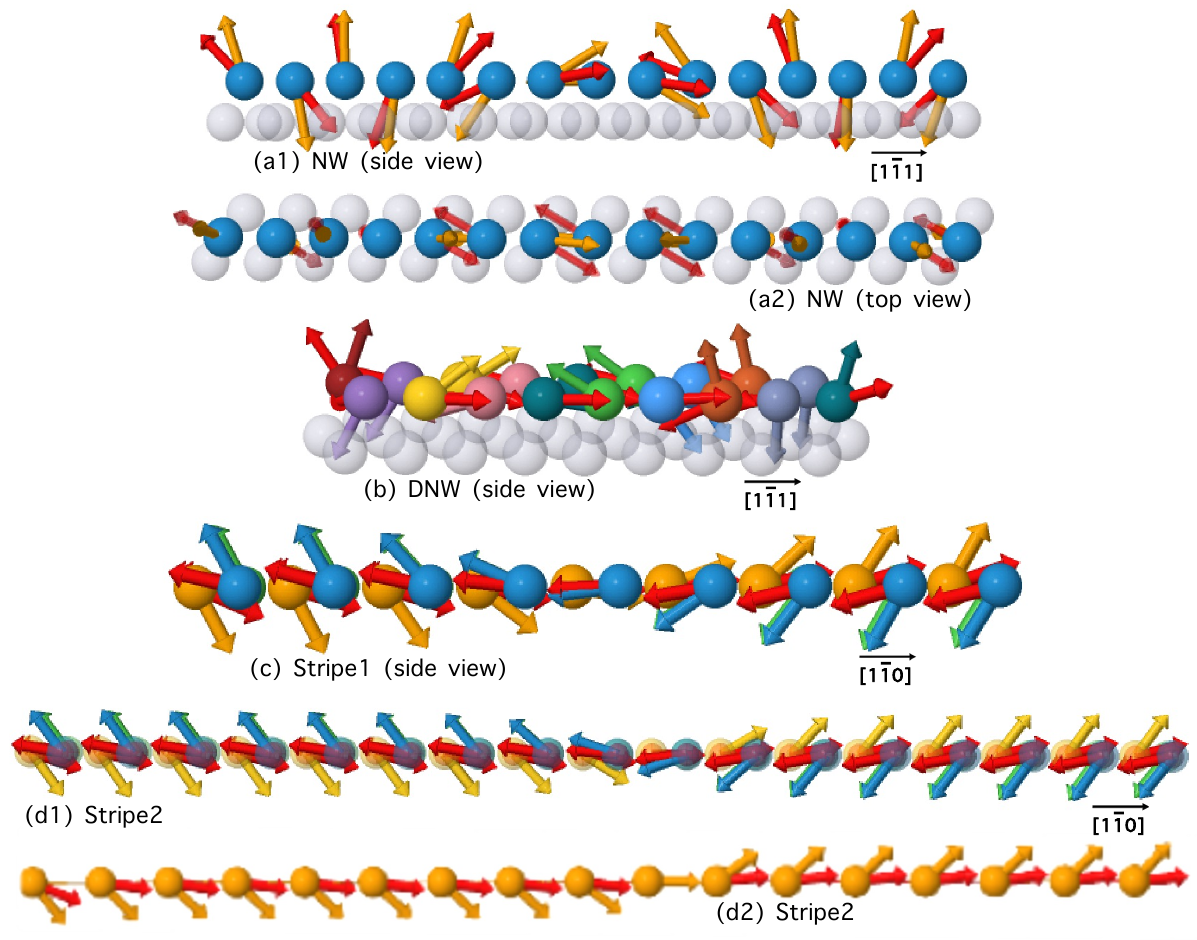}
\caption{(Color online) The spin and orbital ground state magnetic orderings calculated for Mn nanostructures adsorbed on a W(110) surface:
(a1,a2) Mn nanowire (NW); (b) double nanowire (DNW), (c) side view of Stripe-1; (d1) Stripe-2, and (d2) moments directions for only the central row of the Stripe-2 system. 
Yellow and blue arrows represent the directions of the local spin magnetic moment and the red arrows symbolize the corresponding orbital magnetic moment orientations. 
}
\label{fig:orbit-conf}
\end{figure}

\begin{figure}[htp]
\centering
\includegraphics[scale=0.5]{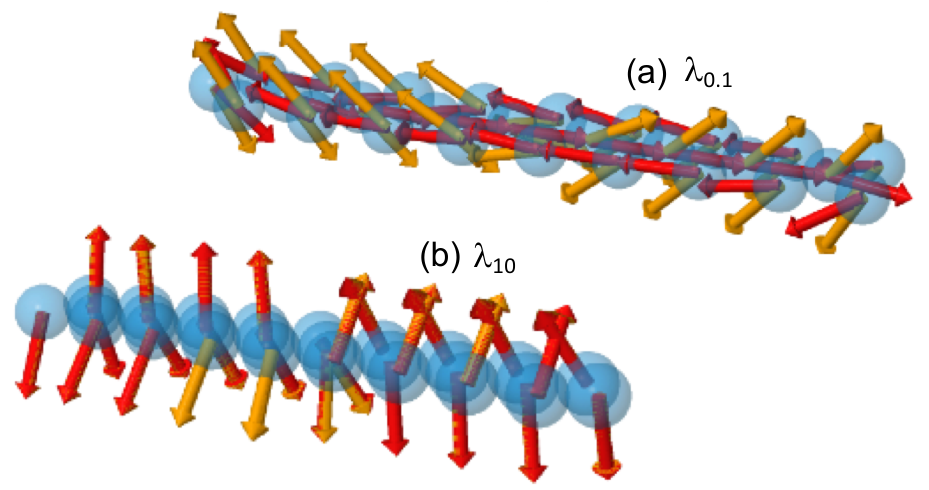}
\caption{(Color online) 
The spin and the orbital magnetic configurations for Mn Stripe-1 on a W(110) surface for different spin-orbit coupling values $\lambda_{\eta}$, were $\eta=1$ refers to 100\% of the true SOC value (Fig. 4(c)), and {\it e.g.} $\eta=0.1$ one tenth of this value.
(a) Stripe-1 for $\eta=0.1$ and (b) for $\eta=10$.
Yellow (red) arrows denote the spin (orbital) moment  directions.
}
\label{fig:fig-soc}
\end{figure}

%
Let us now turn our attention to the ground state orbital magnetic moment configurations of these nanostructures. The calculated orbital moments are depicted by red arrows in Fig.~\ref{fig:orbit-conf}.
The magnitude of the Mn orbital magnetic moments is $\sim$ 0.10 $\mu_\text{B}$/atom, 
whereas for the W atoms it is much smaller $\sim$ 0.005 $\mu_\text{B}$/atom. 
We note that the local spin and orbital magnetic moments of each Mn atom in general are not collinear, as Fig.~\ref{fig:orbit-conf} illustrates (for details see Fig.~\ref{fig:angle-ms-mo-nw} and \ref{fig:colormap-angnw} of the appendix). 
For the NW the orbital and spin moments are at an angle on each Mn site, but they present very similar spiral patterns.
However, in the DNW and the stripes, the orbital moments have a distinct magnetic configuration, which seems rather different from 
the spin arrangement. Except for the edge atoms, the orbital magnetic moments tend to align in-plane along the DNW's length direction, and the same happens for Stripe-1 and Stripe-2 nanostructures (see Fig.~\ref{fig:orbit-conf} and \ref{fig:angle-ms-mo-nw}).
It is instructive to examine the role played by the spin-orbit coupling in the ground-state magnetic configurations of these nanostructures, and to do so we examine how they vary for different values of the SOC strength $\lambda$.
%
Fig.~\ref{fig:fig-soc} shows the results of spin and orbital magnetic moment arrangements calculated self-consistently for Stripe-1 with different values of $\lambda_{\eta}$. Here, $\eta$ is an overall scaling factor: $\lambda_{\eta} = \eta \lambda$, where $\lambda$ represents the values of the SOC intensities calculated for each Mn and W atoms that were used  to obtain the results depicted in Fig.~\ref{fig:orbit-conf}. 

%
We note that the SOC intensity significantly influences the orbital magnetic ordering, as well as the magnitude of the Mn orbital moments, which tends to increase with the enhancement of the SOC coupling values.
For relatively small values of $\lambda_{\eta}$ the spin and orbital moments of each Mn atom are misaligned (see Fig. 5(a)), whereas for sufficiently large SOC values they practically line up (Fig. 5(b)).
%

To better understand the obtained orbital and spin magnetic moment configurations, one may use the following simplified effective Hamiltonian model:
\begin{eqnarray}
\hat H = - \sum_{i \ne j} J_{ij} \hat{ \vec{S_i}} \cdot \hat{ \vec{S_j}} - \lambda \sum_{i} \hat{ \vec{L_i}} \cdot \hat{ \vec {S_i}} - \hat H^{(i)}_{\mathrm{CF}}, 
\label{eq-model}
\end{eqnarray}
where $J_{ij}$ denote the effective Heisenberg (isotropic) exchange interactions between magnetic moments located on sites $i$ and $j$, $\hat{ \vec{S_i}}$ and $\hat{ \vec{L_i}}$ represent the spin and orbital angular momentum operators, respectively, and $\lambda$ is the spin-orbit coupling constant. $\hat H^{(i)}_{\mathrm{CF}}$ symbolizes the crystal field Hamiltonian. 

From the point of view of a more general approach \cite{SECCHI201561} this means that it neglects inter-site orbital orbital and spin orbital interactions, keeping only the onsite contributions.The onsite orbital-orbital interaction is nothing but the crystal field splitting term, as the Stevens operator-equivalent method in crystal-field theory explicates \cite{Stevens}. Neglecting the {\it inter-site} orbital interactions here is justified by the smallness of the orbital magnetic moments.

For simplicity, we shall also neglect the explicit presence of the substrate and deal with an ideal atomic wire, characterized by $C_{\infty v}$ point group symmetry. The leading term in the expansion of the crystal field potential for $C_{\infty v}$ symmetry has the form (see e.g.\cite{MULAK200053}):
\begin{eqnarray}
\hat H^{(i)}_{\mathrm{CF}} = V_{\mathrm{CF}} (3\hat L^2_{i,z}- \hat{\vec {L_i}}^2).
\label{eq-CF}
\end{eqnarray}
Here, we assume that the wire is in $z$ direction. 
Depending on the sign of $V_{\mathrm{CF}}$, the wire will have easy axis or easy plane for orbital magnetic moments.

According to this simplified model, the long-range order character of the spin interactions is primarily defined by the values of the interatomic exchange constants $J_{ij}$. In the Mn chains, for example, the ground state spin spiral-like configuration results from competing interatomic interactions. At the same time, the orbital moment on each atom experiences a combined effect of spin-orbit coupling, which tends to align it parallel to the spin moment, and the crystal field that tends to line it up along the orbital easy-axis (or easy plane) \cite{Laan_2001}. Thus, the final direction of the local orbital angular moments $\vec L$ depends upon the relative strength of these two terms.
In what follows we shall show that this simple model captures the main effects observed in our non-collinear DFT calculations.

We apply this model to a finite-size chain of 20 Mn atoms, treating both spin and orbital moments as classical vectors of unitary length. Since we are dealing with a $3d$ element, we may assume that the $J_{ij}$'s are much larger than $\lambda$. Thus, the ground-state spin configuration is primarily defined by the exchange interactions. Restricting the number of inter-site interactions to nearest ($J_1$) and next-nearest neighbours ($J_2$) only, the ground state spin configuration is defined by the maximum of the Fourier transformed exchange interaction:
\begin{eqnarray}
J(q)=2J_1\cos(q_z a)+2J_2\cos(2q_z a),
\end{eqnarray}
where $q_z$ is the wave-vector component of the spiral and $a$ is the lattice constant. 

\begin{figure*}[!t]
\includegraphics[width=1.8\columnwidth]{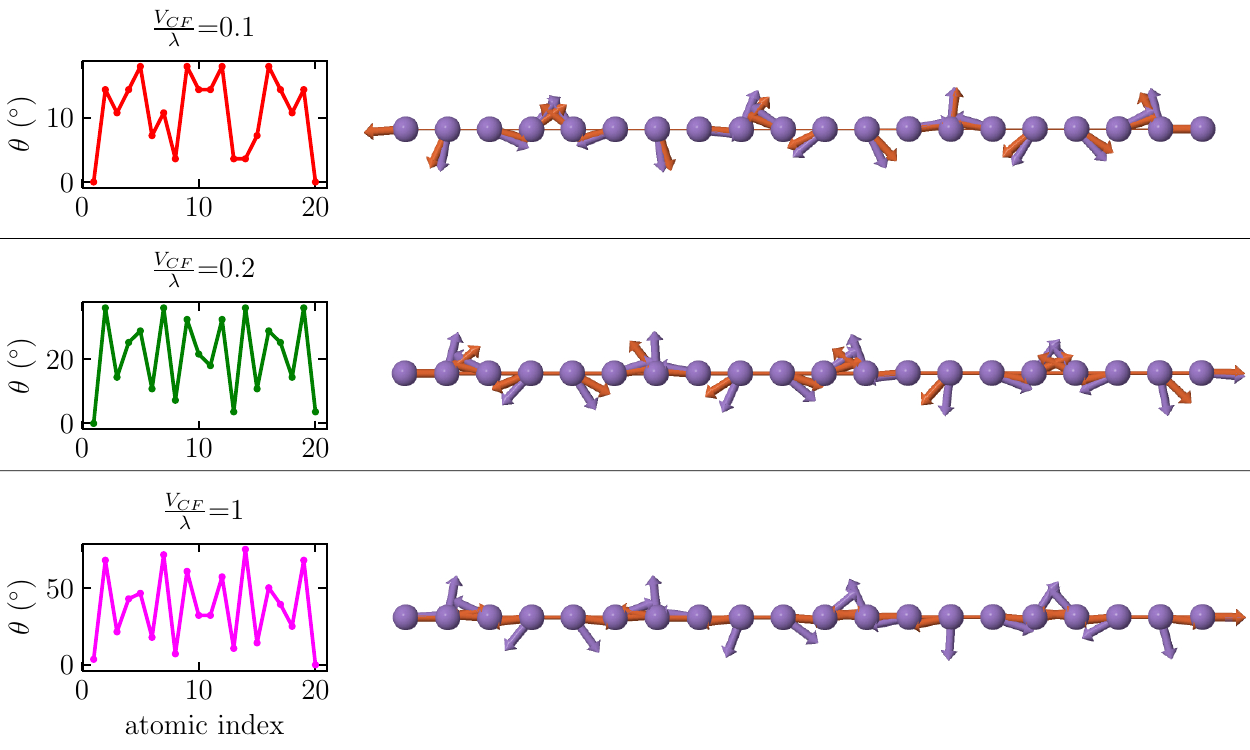} 
\caption{Left: The angles ($\theta_i$, defined in the text) between spin and orbital moments at each site resulting from the model calculation. Note the difference in scale on the y-axis. Right: The ground state magnetic order. Spin (orbital) moments are depicted by violet (orange) arrows. The $z$ axis is along the direction of the wire.}
\label{fig-model}
\end{figure*}

In the model calculations, we set $J_1$=1 (ferromagnetic) and $J_2$=-1 (antiferromagnetic) that leads to a cycloidal spin-spiral like ground state, 
which for an infinite chain has a wavevector $q_{\mathrm{GS}}$=0.21$\times\frac{2\pi}{a}$). 
We assume that both $V_{\mathrm{CF}}$ and $\lambda$ are positive and we vary the ratio between them. The results are shown in Fig.~\ref{fig-model}.
Clearly, when spin-orbit coupling dominates over crystal field effects, i.e., when $V_{\mathrm{CF}} \ll \lambda$, the spin and orbital moments are nearly parallel to each other. This information is shown both in the real space plot of the spin structure (right side of Fig.~\ref{fig-model}) and in the plot of angles between spin and orbital moments (left side of Fig.~\ref{fig-model}). 
As the ratio $r = V_{\mathrm{CF}}/\lambda$ increases, the angles between the spin and orbital moments directions ($\theta_i=\angle (\vec L_i,\vec S_i$)) become bigger, and the orbital moments tend to align more and more along the direction of the chain (which is the easy-axis direction  
for the orbital magnetization). This is illustrated in the middle panel of Fig.~\ref{fig-model} for $r=0.2$. 
Finally, for large values of $r$, the orbital moments align closely with the chain axis, as expected for relatively large values of $V_{\mathrm{CF}}$. The $L^z_i$ components exhibit the same sign as the spin components, while the exchange interactions are responsible for the spin spiral texture.

%
It is worth recalling that $V_{\mathrm{CF}}$ increases with the coordination number. Thus, the crystal field experienced by the central atoms in the ribbons is stronger than that in the chain, while the spin-orbit coupling remains nearly constant. 
This simplified model therefore helps to clarify the mechanisms underlying the results of our non collinear DFT calculations. The combined effect of the spin-orbit coupling and the crystal field favors the alignment of the orbital and spin moments in the NW. In contrast, the central atoms in the ribbons have more neighbors and experience a stronger crystal field, which promotes the alignment of their orbital moments along the easy-axis direction. Furthermore, due to the spin spiral fostered by competing exchange interactions, the orbital moments also rotate around the chain axis, with relatively small transverse components, commensurate with the spin arrangement.

%

\section{Conclusion}
We have conducted first-principles electronic structure calculations to investigate the complex configurations of spin and orbital angular momentum in Mn chains of varying widths (ranging from 1 to 3 atoms) adsorbed on a W(110) surface. We found that the ground-state spin and orbital angular momentum configurations of these nanostructures exhibit noncollinear chiral ordering. Our findings were discussed in light of a simplified model that takes into account Heisenberg effective exchange interactions, spin-orbit coupling, and crystal field effects. The spin arrangements are primarily determined by the competing effective exchange interactions between the first- and second-nearest neighboring Mn atoms. The orbital magnetism is also influenced by the crystal field and spin-orbit interaction, and the balance between them may lead to relatively large tilting angles between the orbital and spin local moments, as we have obtained. Similar features have previously been reported for bulk materials in \onlinecite{PhysRevB.89.064428,PhysRevB.90.165130,PhysRevLett.80.5758}.
%

\section*{Acknowledgements}
M.M.B-N., R.B.M., R.C. and A.B.K. acknowledge financial support from CAPES and CNPq, Brazil. 
A.B.K. acknowledges support from FAPESPA, the INCT of Materials Informatics and the INCT of Spintronics and Advanced Magnetic Nanostructures, CNPq, Brazil.
R.B.M. also acknowledges the INCT of Spintronics and Advanced Magnetic Nanostructures, CNPq, Brazil, R.C. and R.B.M. acknowledge  financial support from FAPERJ - Fundação Carlos Chagas Filho de Amparo à Pesquisa do Estado do Rio de Janeiro, grant number E-26/205.956/2022 and 205.957/2022 (282056).
The calculations were performed at the computational facilities of the CCAD/UFPA (Brazil), National Laboratory for Scientific Computing (SDumont, LNCC/MCTI) and CENAPAD, Brazil. 
O.E. acknowledges support from the Swedish Research Council (VR), the Knut and Alice Wallenberg Foundation, the Swedish Foundation for Strategic Research (SSF), the Swedish Energy Agency (Energimyndigheten), ERC (synergy grant FASTCORR, project 854843), eSSENCE, and STandUPP. M.I.K. acknowledges support from ERC (synergy grant FASTCORR, project 854843) and from the Wallenberg Initiative Materials
Science (WISE), funded by the Knut and Alice Wallenberg
Foundation. The work of Y.K. is supported by VR (project No. 2019-03569) and G{\"o}ran Gustafsson Foundation.


\appendix

\section{Appendix A: Details of the magnetic properties}
 \setcounter{table}{0}
        \renewcommand{\thetable}{A\arabic{table}}%
        \setcounter{figure}{0}
        \renewcommand{\thefigure}{A\arabic{figure}}%
        \renewcommand{\theequation}{A.\arabic{equation}}%
        \setcounter{equation}{0}

\begin{figure}[htp]
\centering
\includegraphics[scale=0.73]{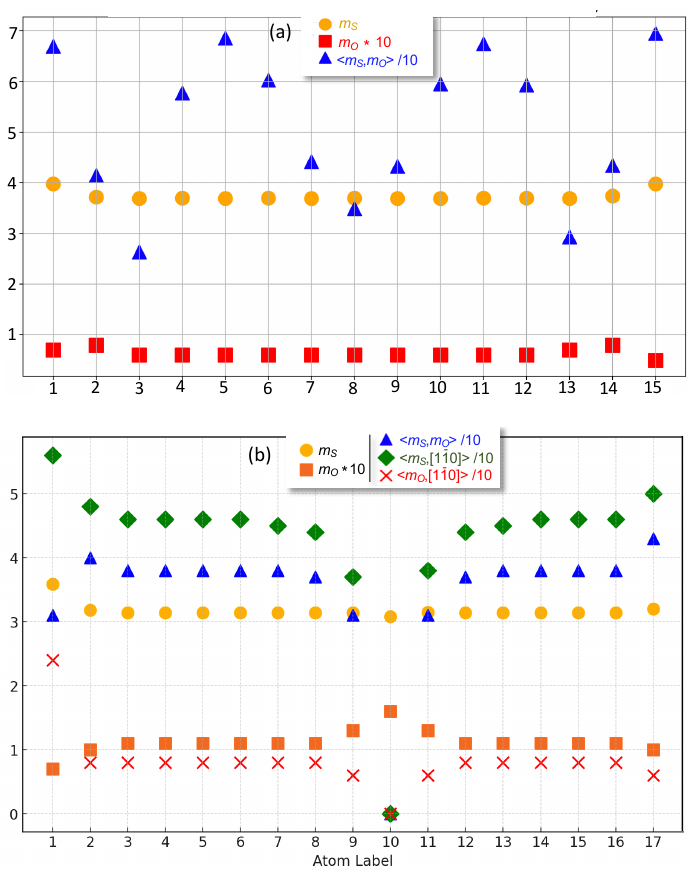}
\caption{ 
Local spin ($m_{S}$) and orbital 
($m_{O}$, multiplied by 10) magnetic moments (in $\mu_{B}$); and absolute values of the angles (in degree) between spin and the orbital moments ($<m_{S},m_{O}>$, divided by 10) at the same Mn atom for the clusters displayed in Fig.~\ref{fig:orbit-conf}(a1,a2) (NW-top panel (a)) and (d1,d2) (Stripe-2-down panel (b)). The downmost numbers (x-axis(atom label)) in the figures refer to the numbering of the atoms in the cluster in Fig.~\ref{fig:systems}.
 For the Stripe-2, Fig.~\ref{fig:orbit-conf} (d2), we present only the values for the atoms located in the central wire, and $<m_{S},[1\overline{1}0]>$ ($<m_{O},[1\overline{1}0]>$) denotes the angle (in degree) between the spin (orbital) moment of each Mn atom and the direction $[1\overline{1}0]$ (both divided by 10).
}
\label{fig:angle-ms-mo-nw}
\end{figure}

\begin{figure}[htp]
\centering
\includegraphics[width=0.9\linewidth]{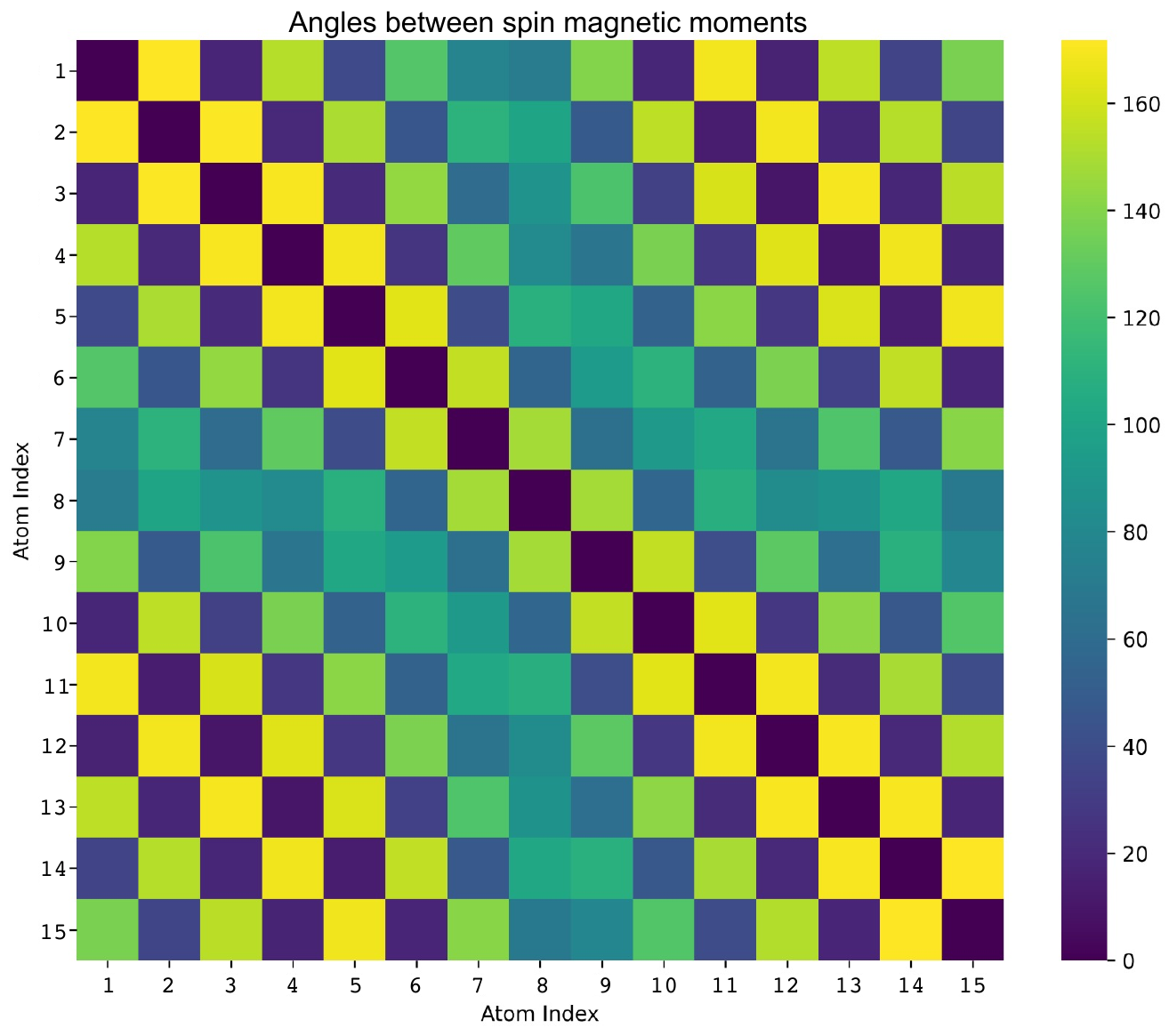}
\\
\includegraphics[width=0.9\linewidth]{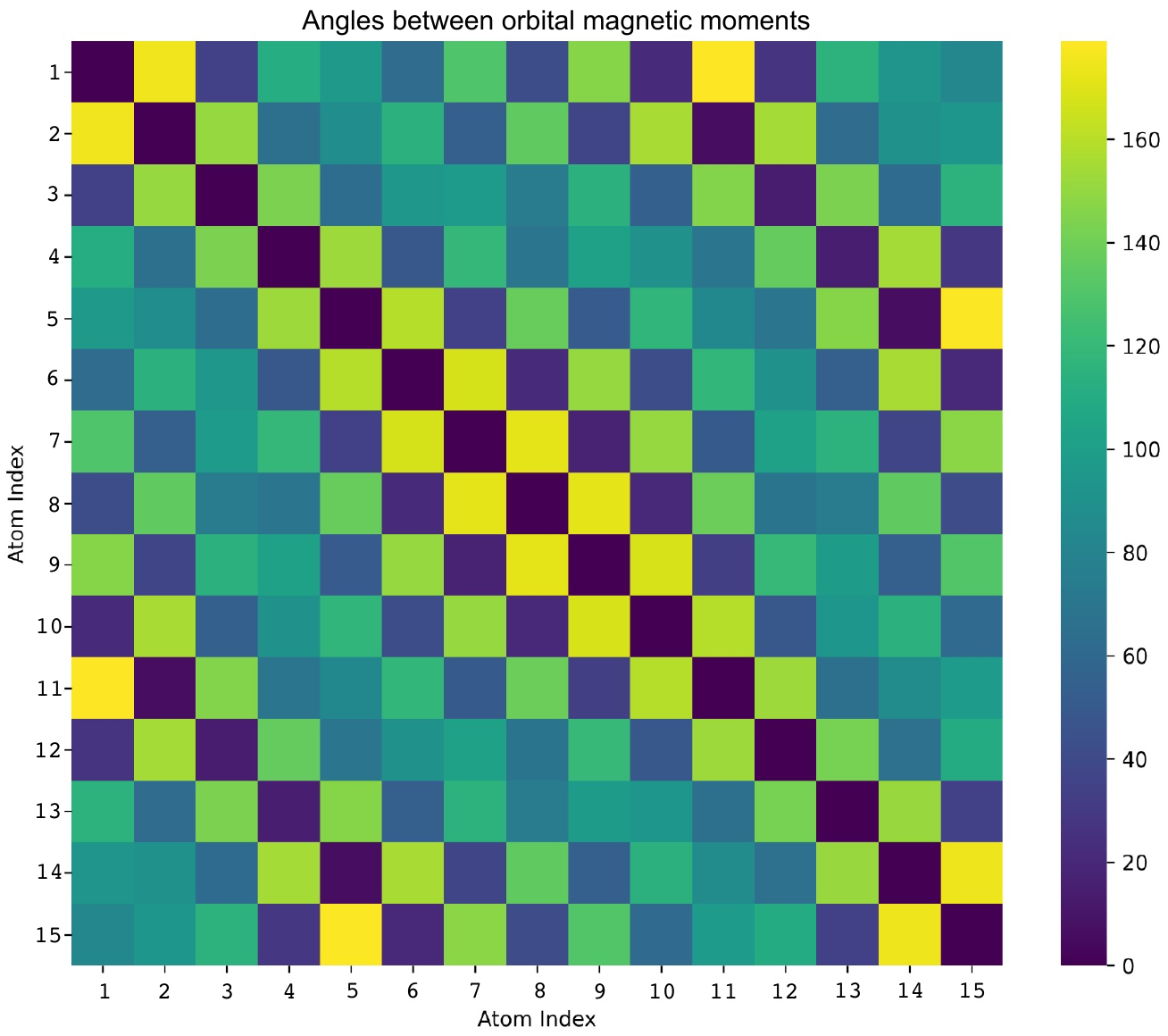}
\caption{\label{fig:heatmap-nw-sp} 
\small{  Angles (in degree) between the Mn local spin ($m_{S}$) (top) and orbital ($m_{O}$) (bottom) moments for the NW magnetic configurations displayed in Fig.~\ref{fig:spin-conf}(a1-a2) and Fig.~\ref{fig:orbit-conf}(a), respectively.
The atom index in the figures refers to the numbering of the atoms in the NW cluster of Fig.~\ref{fig:systems}.
}}
\label{fig:colormap-angnw}
\end{figure}

The spin and orbital magnetic moments and angles between moments for the atoms in the clusters displayed in Fig.~\ref{fig:orbit-conf}(a1,a2) and (d1-d2) are shown 
in Fig.~\ref{fig:angle-ms-mo-nw} and \ref{fig:colormap-angnw}, 
where the atoms are labeled according to Fig.~\ref{fig:systems}. 
The angles between the spin and orbital moments at the same site ($<m_{S},m_{O}>$) are also shown in 
Fig.~\ref{fig:angle-ms-mo-nw}.

\begin{figure}[htp]
\centering
\includegraphics[width=0.99\linewidth]{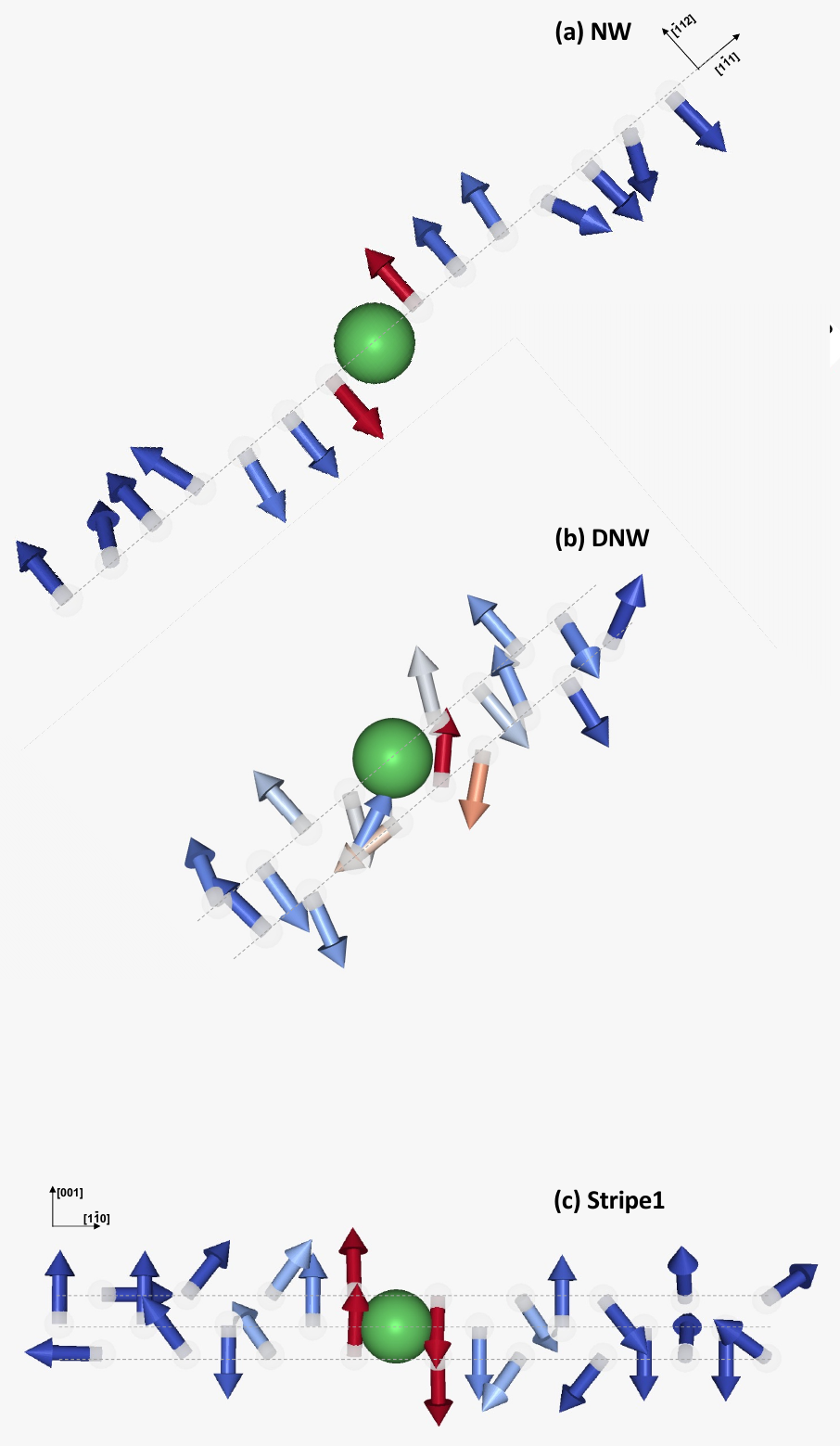}
\caption{\small{DMI directions between a Mn central  atom (green circle) and the Mn atom in the site of the arrow in (a) Mn NW, (b) DNW and (c) Stripe-1 on W(110).
}}
\label{fig:dm-direction}
\end{figure}

The DM directions between a central Mn site and its nearest neighbors sites for the NW and  the Stripe-2 are presented in Fig.~\ref{fig:dm-direction}.

\clearpage
%

\bibliographystyle{apsrev4-1}


\end{document}